\documentclass[12pt]{article}
\usepackage{multicol}
\usepackage{amssymb}
\usepackage{amsfonts,amssymb,amsmath,amsthm}
\usepackage{epsfig}
\usepackage{slashed}
\usepackage{graphicx}
\usepackage{subfig}
\usepackage{mathrsfs}
\usepackage{bbm}
\usepackage{cite}
\usepackage{enumerate}
\usepackage{slashbox}
\usepackage{color}
\usepackage[colorlinks,linkcolor=red,anchorcolor=red,citecolor=green]{hyperref}
\makeatletter 
\@addtoreset{equation}{section}
\makeatother  

\begin{document}
\renewcommand{\thefootnote}{\fnsymbol{footnote}}
\begin{titlepage}

\vspace{10mm}
\begin{center}
{\Large\bf Thermodynamics of Horndeski black holes with non-minimal derivative coupling}
\vspace{16mm}

{{\large Yan-Gang Miao${}^{1,2,}$\footnote{\em E-mail: miaoyg@nankai.edu.cn}
and Zhen-Ming Xu}${}^{1,}$\footnote{\em E-mail: xuzhenm@mail.nankai.edu.cn}

\vspace{6mm}
${}^{1}${\normalsize \em School of Physics, Nankai University, Tianjin 300071, China}

\vspace{3mm}
${}^{2}${\normalsize \em Max-Planck-Institut f\"ur Gravitationsphysik (Albert-Einstein-Institut),\\
M\"uhlenberg 1, D-14476 Potsdam, Germany}

}

\end{center}

\vspace{10mm}
\centerline{{\bf{Abstract}}}
\vspace{6mm}
We explore thermodynamic properties of a new class of Horndeski black holes whose action contains a non-minimal kinetic coupling of a massless real scalar and the Einstein tensor. Our treatment is based on the well-accepted consideration, where the cosmological constant is dealt with as thermodynamic pressure and the mass of  black holes as thermodynamic enthalpy. We resort to a newly introduced intensive thermodynamic variable, i.e., the coupling strength of the scalar and tensor whose dimension is length square, and thus yield both the generalized first law of thermodynamics and the generalized Smarr relation. Our result indicates that this class of Horndeski black holes presents rich thermodynamic behaviors and critical phenomena. Especially in the case of the presence of an electric field, these black holes undergo two phase transitions. Once the charge parameter exceeds its critical value, or the cosmological parameter does not exceed its critical value, no phase transitions happen and the black holes are stable. As a by-product, we point out that the coupling strength acts as the thermodynamic pressure in the behavior of thermodynamics.
\vskip 20pt
\noindent
{\bf PACS Number(s)}: 04.50.Kd, 04.70.Dy, 04.70.Bw


\end{titlepage}

\newpage
\renewcommand{\thefootnote}{\arabic{footnote}}
\setcounter{footnote}{0}
\setcounter{page}{2}
\pagenumbering{arabic}
\tableofcontents
\vspace{1cm}

\section{Introduction}
General relativity and quantum mechanics have become two pillars of modern physics. Meanwhile, due to constant developments of technologies, increasingly accurate observations mostly indicate that the Einstein's theory of general relativity has passed all experimental tests with flying colors, especially in the weak-field or slow-motion regime~\cite{EB}. More attractively, black holes, which can probably be regarded as a tie of connecting general relativity and quantum mechanics~\cite{XC}, have been getting more and more attentions. In particular, the thermodynamics of black holes in anti-de Sitter (AdS) spacetime~\cite{JMB,RW,SC} has acquired great progress since the AdS/CFT duality plays a pivotal role in recent developments of theoretical physics~\cite{SH,RGC}.

Conversely, some of the funniest strong-field predictions of the Einstein's theory of general relativity still remain difficult to be understood and verified. In this sense, black holes are ideal candidates to be used as probes of Einstein theory. Most theoretical and observational issues, such as the high-curvature corrections, the origin of curvature singularities, the cosmological constant problem, the dark energy/matter, and so on, strongly recommend that the Einstein's theory of general relativity should be modified, i.e., the so-called modified gravity~\cite{TC,EP}. What is worth mentioning is that Horndeski~\cite{GWH} proposed the most general scalar-tensor modified gravity action which generates equations of motion with second-order derivatives. The Horndeski scalar-tensor modified gravity theory has widely been investigated in astrophysics and cosmology~\cite{ESSVS,CCEJA,AMMHO}. In the investigation of a locally stable solution of Horndeski black holes, the action containing a non-minimal kinetic coupling of one scalar and Einstein tensor has received considerable attentions, and some important progress has been made. A spherically symmetric and static solution has been obtained in ref.~\cite{MR} for the case of a vanishing cosmological constant, and in refs.~\cite{AAJ,MM} for the case of a negative cosmological constant. The no-hair theorem for scalar-tensor gravity theory has been shown in refs.~\cite{LHAN,TPS}. Moreover, the black hole solution in the presence of an electric field~\cite{ACCE,TK,XHHC}, the BTZ black hole solution with a Horndeski source~\cite{MBMH}, and the slowly rotating black hole solutions~\cite{AHME} have been studied. Furthermore, the black hole solution with a time-dependent scalar~\cite{EBCC,MBMH1} and the exact wormhole solutions with a non-minimal kinetic coupling~\cite{RSVS} have also been found. For the topics in other relevant aspects, see, for instance, refs.~\cite{EBKY,HTT,XHHC1,TKNT,RMLLJ,ECCAL,MR2}.

In this paper we revisit thermodynamic properties of Horndeski black holes with a non-minimal kinetic coupling in the presence of an electric field along the line of refs.~\cite{CEJM,DSJT,BPD2,BPD,KM}, namely by considering the cosmological constant as thermodynamic pressure and the mass of black holes as thermodynamic enthalpy. A few thermodynamic quantities have been calculated~\cite{MR,AAJ,MM} for Horndeski black holes, showing that the first law of thermodynamics is satisfied, but the Smarr relation is violated. Hence, we wish to fill up this deficiency from the point of view of thermodynamics. If we take the non-minimal kinetic coupling strength of scalar and tensor fields into account, the relevant term should appear in the Smarr relation and its variation should be included in the first law of thermodynamics. As a result, we obtain the generalized Smarr relation and the first law of thermodynamics in the extended phase space that includes the coupling strength and its conjugate. In other words, we shall deal with thermodynamic behaviors of Horndeski black holes in a new extended phase space. It has been known that the Born-Infeld parameter~\cite{GKM,NB}, the Gauss-Bonnet coupling constant~\cite{RCLY}, and the noncommutative parameter~\cite{YGMX} can be dealt with as a kind of thermodynamic pressure.  
Our result further indicates that the coupling strength acts as the thermodynamic pressure in the behavior of thermodynamics. Meanwhile, we show that the class of Horndeski black holes with a non-minimal kinetic coupling presents rich critical phenomena.

The paper is organized as follows. In section \ref{sec2}, the thermodynamics of Horndeski black holes with a non-minimal derivative coupling is analyzed. This section contains two subsections which correspond to the scenarios without and with charge, respectively. Finally, we devote to drawing our conclusion in section \ref{sec4}. In addition, the geometric units, $\hbar=c=k_{B}=G=1$, are adopted throughout this paper.

\section{Thermodynamics of Horndeski black holes}\label{sec2}
At the beginning, we proceed to investigate the class of Horndeski black holes whose action contains a non-minimal kinetic coupling of the massless real scalar $\phi$ and the Einstein tensor $G_{\mu\nu}$. This action, in the presence of an electric field, has the following form~\cite{ACCE,XHHC},
\begin{equation}
I=\int \sqrt{-g} \,\text{d}^4 x \left[(R-2\Lambda)-\frac12(\alpha g_{\mu\nu}-\eta G_{\mu\nu})\nabla^{\mu}\phi \nabla^{\nu}\phi-\frac14 F_{\mu\nu}F^{\mu\nu}\right], \label{act}
\end{equation}
where $\eta$ stands for the coupling strength with the dimension of length square, $\alpha$ a coupling constant, $\Lambda$ the negative cosmological constant, $R$ the scalar curvature, $g_{\mu\nu}$ the metric with mostly plus signatures, and $F_{\mu\nu}$ the electromagnetic field strength defined as $F_{\mu\nu} \equiv \partial_{\mu}A_{\nu}-\partial_{\nu}A_{\mu}$ with $A_{\mu}$ the vector potential.

Making a variation of the action eq.~(\ref{act}) with respect to the metric $g_{\mu\nu}$, the scalar field $\phi$, and the Maxwell field $A_{\mu}$, respectively, one can obtain
\begin{eqnarray}
G_{\mu\nu}+\Lambda g_{\mu\nu}&=&\frac12 \left(\alpha T_{\mu\nu}+\eta \Xi_{\mu\nu}+E_{\mu\nu}\right), \label{gmu}\\
\nabla_{\mu}\left[(\alpha g_{\mu\nu}-\eta G_{\mu\nu})\nabla_{\nu}\phi\right]&=&0, \label{scalarmu}\\
\nabla_{\mu}F^{\mu\nu}&=&0, \label{emu}
\end{eqnarray}
where $T_{\mu\nu}$, $\Xi_{\mu\nu}$, and $E_{\mu\nu}$ are defined as
\begin{eqnarray}
T_{\mu\nu} &\equiv& \nabla_{\mu}\phi \nabla_{\nu}\phi-\frac12 g_{\mu\nu}\nabla_{\rho}\phi \nabla^{\rho}\phi,\\
\Xi_{\mu\nu} &\equiv & \frac12 \nabla_{\mu}\phi \nabla_{\nu}\phi R-2\nabla_{\rho}\phi \nabla_{(\mu}\phi R_{\nu)}^{\rho}-\nabla^{\rho}\phi \nabla^{\lambda}\phi R_{\mu\rho\nu\lambda}\nonumber \\
& &-(\nabla_{\mu}\nabla^{\rho}\phi)(\nabla_{\nu}\nabla_{\rho}\phi)
+(\nabla_{\mu}\nabla_{\nu}\phi)\square \phi+\frac12 G_{\mu\nu}(\nabla \phi)^2\nonumber\\
& &-g_{\mu\nu}\left[-\frac12 (\nabla^{\rho}\nabla^{\lambda}\phi)(\nabla_{\rho}\nabla_{\lambda}\phi)
+\frac12 (\square \phi)^2-\nabla_{\rho}\phi \nabla_{\lambda}\phi R^{\rho\lambda}\right],\\
E_{\mu\nu}&\equiv& F_{\mu}^{\rho}F_{\nu\rho}-\frac12 g_{\mu\nu}F^2.
\end{eqnarray}

In the following we focus on the static solutions with the spherical symmetry in eqs.~(\ref{act})-(\ref{emu}), so the metric is simplified to be
\begin{eqnarray}
\text{d} s^2&=&-f(r)\text{d} t^2+g(r)\text{d} r^2+r^2(\text{d} \theta^2+\sin^2 \theta \text{d}\varphi^2), \label{metric}\\
F&=&\text{d}A,\\
A&=&\Psi \text{d}t,
\end{eqnarray}
where $f(r)$ and $g(r)$ are functions to be determined and $\Psi$ is the electrostatic potential. Under the assumption of spherical symmetry, we only consider a static and isotropic scalar field, i.e., the scalar field is a function of the radial coordinate, $\phi=\phi(r)$. In the two subsections below we shall discuss thermodynamic properties of such a class of Horndeski black holes without and with the Maxwell field, respectively, i.e., under the considerations of the specific forms of $f(r)$, $g(r)$, and $\phi(r)$ in the former subsection, and the specific forms of $f(r)$, $g(r)$, $\phi(r)$, and $\Psi(r)$ in the latter one.

\subsection{Scenario without charge}
For this situation, the analytic solution takes the form~\cite{AAJ,MM},
\begin{eqnarray}
f(r)&=&\frac{\alpha r^2}{3\eta}-\frac{2M}{r}+\frac{3\alpha+\Lambda \eta}{\alpha-\Lambda \eta}+\left(\frac{\alpha+\Lambda \eta}{\alpha-\Lambda \eta}\right)^2\,\frac{\tan^{-1}\left(\sqrt{\frac{\alpha}{\eta}}r\right)}{\sqrt{\frac{\alpha}{\eta}}r},\label{ff}\\
g(r)&=&\frac{\alpha^2[(\alpha-\Lambda \eta) r^2+2\eta]^2}{(\alpha-\Lambda \eta)^2(\alpha r^2+\eta)^2 f(r)},\\
\vspace{0.4cm}
\psi^2(r)&=&-\frac{2\alpha^2 r^2(\alpha+\Lambda \eta)[(\alpha-\Lambda \eta) r^2+2\eta]^2}{\eta(\alpha-\Lambda \eta)^2 (\alpha r^2+\eta)^3 f(r)}, \label{scalar1}
\end{eqnarray}
where $M$ is considered as the mass of black holes, and $\psi \equiv \phi^{\prime}$, where a prime stands for the first order derivative with respect to $r$. This solution requires $\alpha$ and $\eta$ to have the same sign and $\alpha \neq -\Lambda \eta$. Once $\alpha=-\Lambda \eta$, the Schwarzschild-AdS solution is recovered and the scalar field becomes trivial~\cite{MM}. For simplicity but without the loss of generality, we set $\alpha >0$ and $\eta >0$ in the following context.

Based on refs.~\cite{AAJ,MM}, we write the constraint\footnote{This constraint condition does not work on the analysis of thermodynamic properties because we adopt the method of horizon thermodynamics. Incidentally, its corresponding form for the scenario with charge is given by eq.~(\ref{yueshu2}).} that ensures the reality of the scalar field outside the horizon,
\begin{equation}
\alpha+\Lambda \eta <0. \label{yueshu1}
\end{equation}
It is obvious to deduce such a condition from eq.~(\ref{ff}) and eq.~(\ref{scalar1}).
At first, let us see the asymptotic behavior of eq.~(\ref{ff}): $f(r)$ goes to minus infinity under the limit $r \rightarrow 0$; on the other hand, it goes to plus infinity under the limit $r \rightarrow +\infty$. Therefore, the equation $f(r)=0$ has at least one real root and the largest real root can be regarded as the horizon radius $r_h$. Next, it is evident that $f(r)>0$ once $r>r_h$. Hence, the positivity of eq.~(\ref{scalar1}), i.e., the reality of $\phi(r)$ in the regime $r>r_h$ leads of course to the above inequality.
In addition, it is necessary to take a close look at the behavior of the scalar field in the near horizon region because the scalar field seems to be divergent from eq.~(\ref{scalar1}). Due to $f(r_h)=0$ and $f^{\prime}(r_h)\neq 0$, one can get the Taylor expansion of $f(r)$:  $f(r)=f_0+f_1(r-r_h)+f_2(r-r_h)^2 +\cdots$. As to $\psi(r)$, see eq.~(\ref{scalar1}), it approximates to $\frac{1}{\sqrt{f(r)}}$ in this region, which gives rise to the form of the scalar field, $\phi(r)=\phi_0+\phi_1(r-r_h)^{1/2}+\phi_2(r-r_h)^{3/2}+\cdots$. As a result, the scalar field remains finite in the near horizon region.

Now let us revisit the thermodynamic properties of this class of Horndeski black holes.
Along the line of refs.~\cite{DSJT,BPD2,BPD,KM}, one can regard the mass of black holes as the thermodynamic enthalpy,
\begin{equation}
M=\frac{\alpha r_h^3}{6\eta}+\frac{(3\alpha+\Lambda \eta)r_h}{2(\alpha-\Lambda \eta)}+\frac12\sqrt{\frac{\eta}{\alpha}}\left(\frac{\alpha+\Lambda \eta}{\alpha-\Lambda \eta}\right)^2 \tan^{-1}\left(\sqrt{\frac{\alpha}{\eta}}r_h\right). \label{enth1}
\end{equation}
The Bekenstein-Hawking entropy is one-fourth of the event horizon area,
\begin{equation}
S =\pi r_h^2, \label{entro}
\end{equation}
and the thermodynamic temperature can be calculated to be
\begin{equation}
T_h=\left(\frac{\partial M}{\partial S}\right)_{\eta,\Lambda}=\frac{1}{4\pi r_h}\left[\frac{\alpha r_h^2}{\eta}+\frac{\eta(\alpha+\Lambda \eta)^2}{(\alpha r_h^2+\eta)(\alpha-\Lambda \eta)^2}+\frac{3\alpha+\Lambda \eta}{\alpha-\Lambda \eta}\right]. \label{temp1}
\end{equation}
When the thermodynamic pressure $P$ is regarded as
\begin{equation}
P=-\frac{\Lambda}{8\pi}, \label{press}
\end{equation}
the extensive variable conjugate to it, i.e., the thermodynamic volume $V$  has the form,
\begin{eqnarray}
V=\left(\frac{\partial M}{\partial P}\right)_{\eta,S}=-\frac{16\pi \alpha \eta}{(\alpha-\Lambda \eta)^3}\left[(\alpha-\Lambda \eta)r_h+\sqrt{\frac{\eta}{\alpha}}(\alpha+\Lambda \eta)\tan^{-1}\left(\sqrt{\frac{\alpha}{\eta}}r_h\right)\right].\label{case1V}
\end{eqnarray}

Next, we introduce a new intensive thermodynamic variable $\Pi$ in terms of the coupling strength $\eta$,
\begin{equation}
\Pi \equiv\frac{\alpha}{8\pi \eta}, \label{new}
\end{equation}
and derive, with the help of eq.~(\ref{enth1}), the extensive variable conjugate to it, 
\begin{eqnarray}
\Theta &=&\left(\frac{\partial M}{\partial \Pi}\right)_{P,S} \nonumber\\
&=&\frac{4\pi r_h^3}{3}+\frac{2\pi \eta^2 r_h(\alpha+\Lambda \eta)^2}{\alpha(\alpha r_h^2+\eta)(\alpha-\Lambda \eta)^2} -2\pi \left(\frac{\eta}{\alpha}\right)^{\frac32}\left(\frac{\alpha+\Lambda \eta}{\alpha-\Lambda \eta}\right)^2\tan^{-1}\left(\sqrt{\frac{\alpha}{\eta}}r_h\right) \nonumber\\
& &-\frac{16\pi\Lambda\eta^2 r_h}{(\alpha-\Lambda\eta)^2}\left[1+\frac{\alpha+\Lambda \eta}{\alpha-\Lambda \eta}\,\frac{\tan^{-1}\left(\sqrt{\frac{\alpha}{\eta}}r_h\right)}{\sqrt{\frac{\alpha}{\eta}}r_h}\right].\label{case1Theta}
\end{eqnarray}
As a result, we can write the first law of thermodynamics,
\begin{equation}
\text{d}M=T_h \text{d}S+V\text{d}P+\Theta \text{d}\Pi, \label{law1}
\end{equation}
and the generalized Smarr relation,
\begin{equation}
M=2T_h S-2PV-2  \Pi \Theta. \label{law2}
\end{equation}
We thus provide a possibility of making up the gap in refs.~\cite{MR,AAJ,MM}, i.e., the Smarr relation can be maintained in the extended phase space that contains $\Pi$ and $\Theta$.

The heat capacity at constant pressure is defined by
\begin{equation}
C_p \equiv \left(\frac{\partial M}{\partial T_h}\right)_P=\frac{\partial M}{\partial r_h}\left(\frac{\partial T_h}{\partial r_h}\right)^{-1}, \label{capac1}
\end{equation}
where the two factors can be calculated to be
\begin{eqnarray*}
\frac{\partial M}{\partial r_h} &=& \frac{\alpha r_h^2}{2\eta}+\frac{\eta(\alpha+\Lambda\eta)^2}{(\alpha r_h^2+\eta)(\alpha-\Lambda\eta)^2}+\frac{3\alpha+\Lambda\eta}{\alpha-\Lambda\eta},\\
\vspace{0.4cm}
\frac{\partial T_h}{\partial r_h}&=&-\frac{1}{4\pi r_h^2}\left[-\frac{\alpha r_h^2}{\eta}+\frac{\eta(\alpha+\Lambda\eta)^2}{(\alpha r_h^2+\eta)(\alpha-\Lambda\eta)^2}+\frac{3\alpha+\Lambda\eta}{\alpha-\Lambda\eta}+\frac{2\alpha\eta r_h^2(\alpha+\Lambda\eta)^2}{(\alpha r_h^2+\eta)^2(\alpha-\Lambda\eta)^2}\right].
\end{eqnarray*}

The Gibbs free energy is the Legendre transform of the enthalpy eq.~(\ref{enth1}), i.e. $G\equiv M-T_h S$. Thanks to eqs.~(\ref{enth1})-(\ref{temp1}), we obtain its exact expression,
\begin{equation}
G=-\frac{\alpha r_h^3}{12\eta}+\frac{(3\alpha+\Lambda \eta)r_h}{4(\alpha-\Lambda \eta)}-\frac{\eta r_h(\alpha+\Lambda \eta)^2}{4(\alpha r_h^2+\eta)(\alpha-\Lambda \eta)^2}+\frac12\sqrt{\frac{\eta}{\alpha}}\left(\frac{\alpha+\Lambda \eta}{\alpha-\Lambda \eta}\right)^2 \tan^{-1}\left(\sqrt{\frac{\alpha}{\eta}}r_h\right).  \label{gfree1}
\end{equation}

\begin{figure}
\begin{center}
  \begin{tabular}{cc}
    \includegraphics[width=70mm]{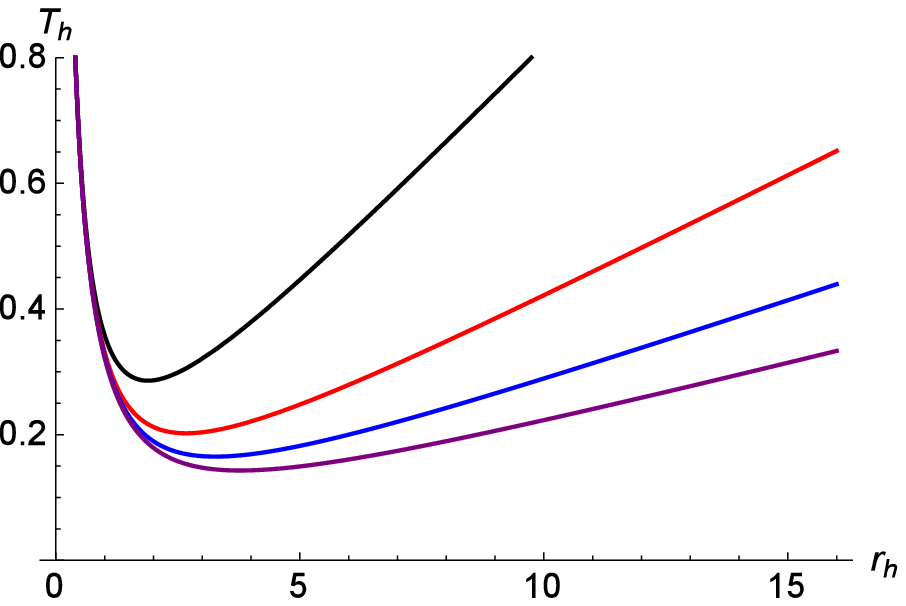} &
    \includegraphics[width=70mm]{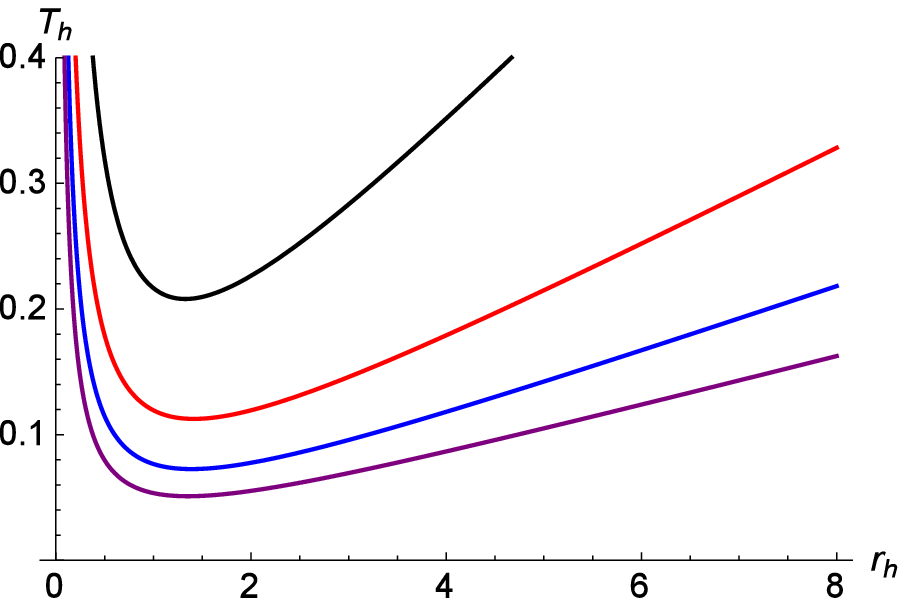} \\
    \includegraphics[width=70mm]{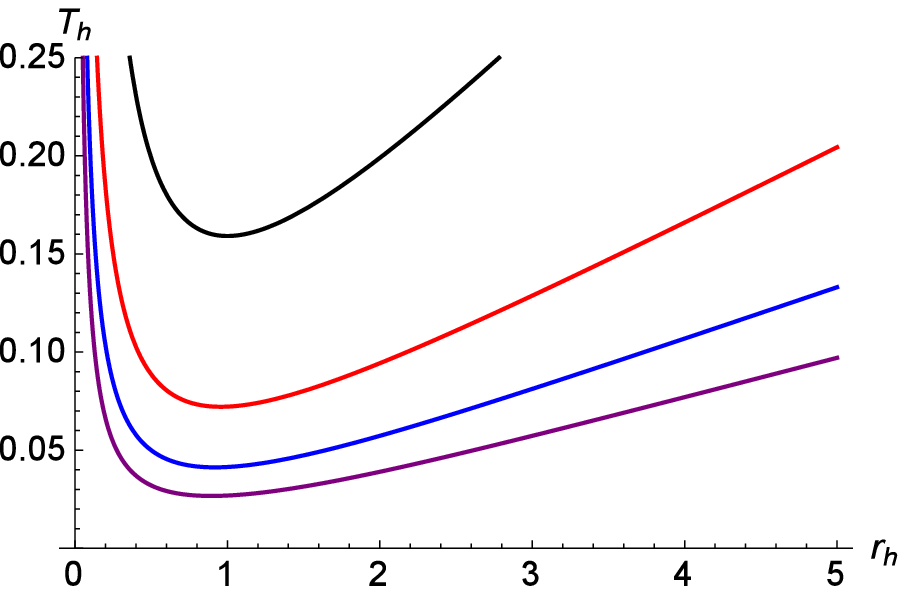} &
    \includegraphics[width=70mm]{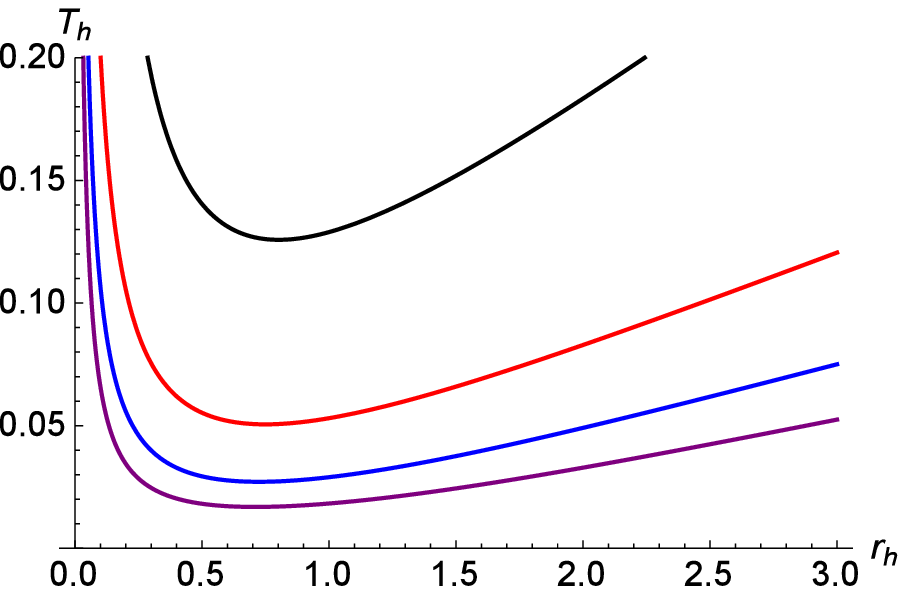}
  \end{tabular}
\end{center}
\caption{When $\alpha=1$, plots of the relation of $T_h$ with respect to $r_h$ for $\eta=1$ (\text{Black}), $2$ (\text{Red}), $3$ (\text{Blue}), and $4$ (\text{Purple}) at $\Lambda=0$ (\text{top left}), $\Lambda=-0.5$ (\text{top right}), $\Lambda=-1$ (\text{bottom left}), and $\Lambda=-1.5$ (\text{bottom right}), respectively.}
\label{tu1}
\end{figure}

\begin{figure}
\begin{center}
  \begin{tabular}{cc}
    \includegraphics[width=70mm]{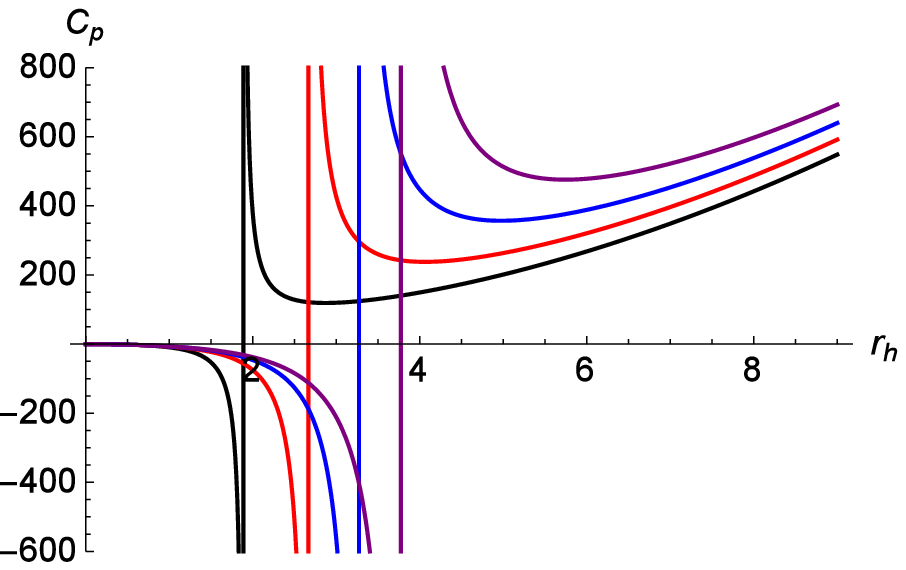} &
    \includegraphics[width=70mm]{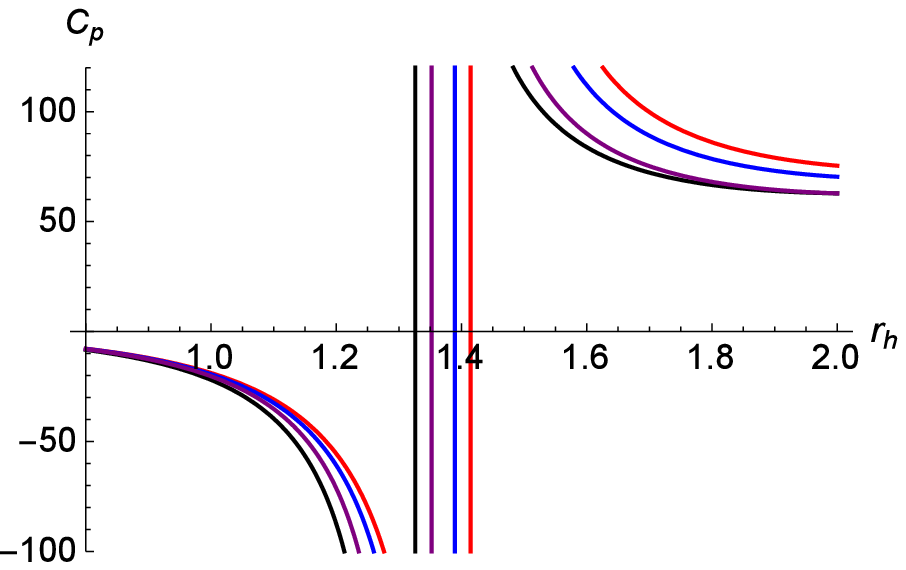} \\
    \includegraphics[width=70mm]{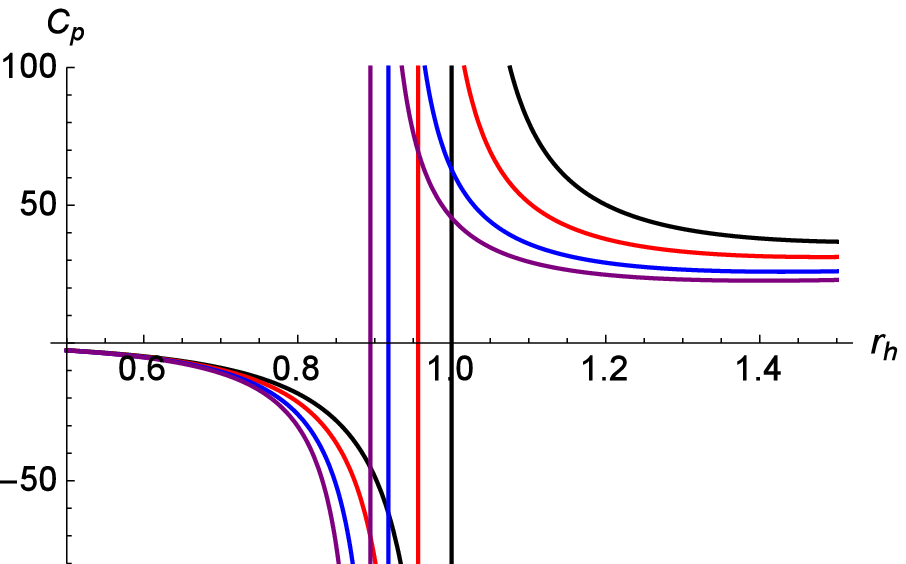} &
    \includegraphics[width=70mm]{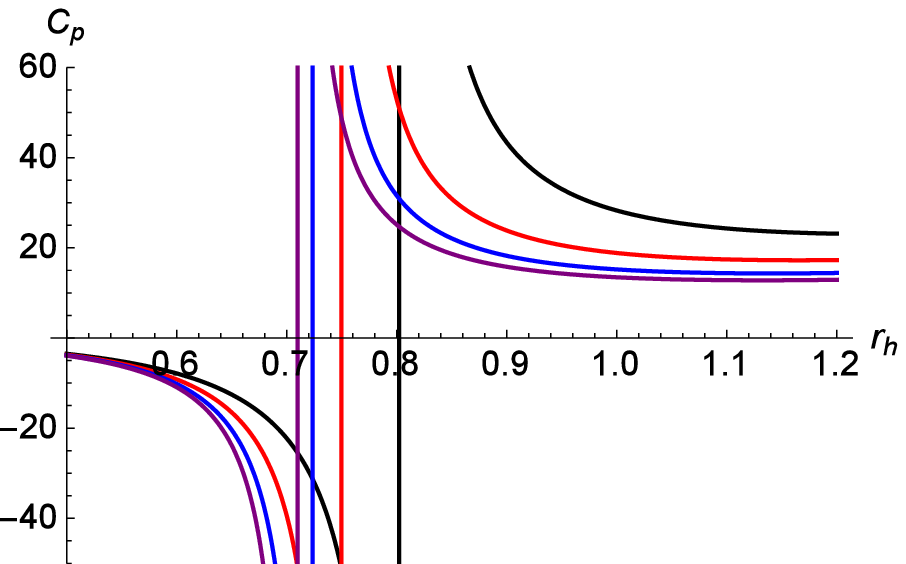}
  \end{tabular}
\end{center}
\caption{When $\alpha=1$, plots of the relation of $C_p$ with respect to $r_h$ for $\eta=1$ (\text{Black}), $2$ (\text{Red}), $3$ (\text{Blue}), and $4$ (\text{Purple}) at $\Lambda=0$ (\text{top left}), $\Lambda=-0.5$ (\text{top right}), $\Lambda=-1$ (\text{bottom left}), and $\Lambda=-1.5$ (\text{bottom right}), respectively.}
\label{tu2}
\end{figure}

\begin{figure}
\begin{center}
  \begin{tabular}{cc}
    \includegraphics[width=70mm]{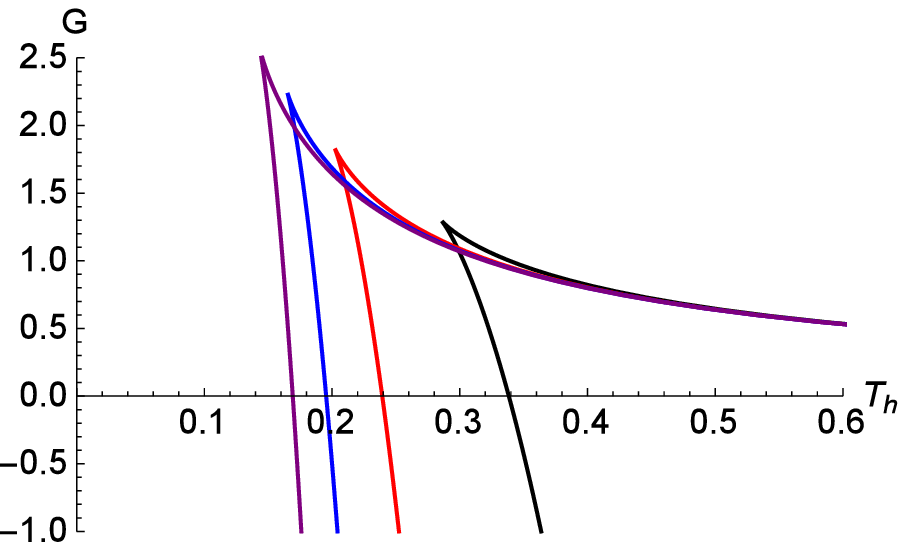} &
    \includegraphics[width=70mm]{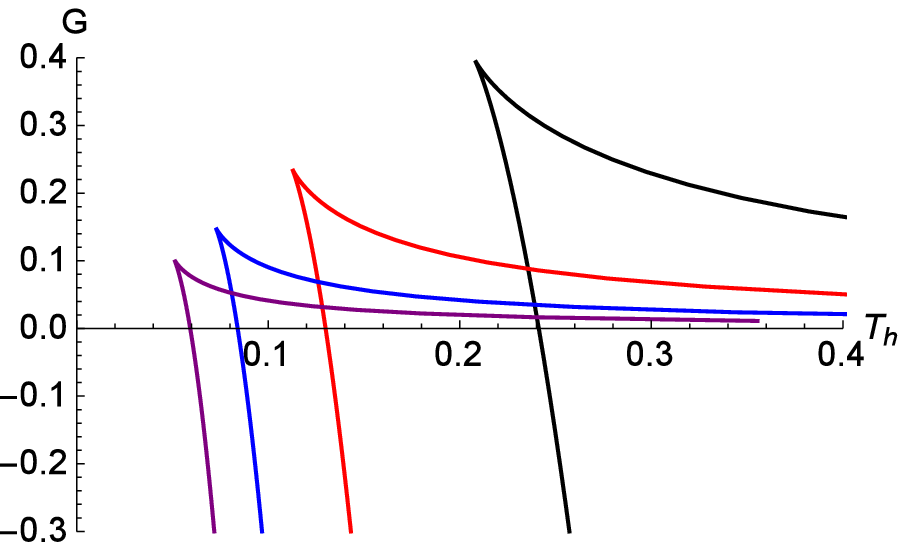}
  \end{tabular}
\end{center}
\caption{When $\alpha=1$, plots of the relation of $G$ with respect to $T_h$ for $\eta=1$ (\text{Black}), $2$ (\text{Red}), $3$ (\text{Blue}), and $4$ (\text{Purple}) at $\Lambda=0$ (\text{left}) and $\Lambda=-0.5$ (\text{right}), respectively.}
\label{tu3}
\end{figure}

In order to visualize the thermodynamic quantities, we plot the temperature eq.~(\ref{temp1}), the heat capacity at constant pressure eq.~(\ref{capac1}), and the Gibbs free energy eq.~(\ref{gfree1}) in Figures \ref{tu1}, \ref{tu2} and \ref{tu3}. From the three figures, we can see that the thermodynamic behaviors are similar for $\Lambda=0$ and $\Lambda\neq 0$. For case $\Lambda=0$, according to eq.~(\ref{scalar1}), the scalar field outside the horizon is not real and it can be explained as an extra degree of freedom, rather than a matter field~\cite{MR}. Fortunately, for $\Lambda\neq 0$, i.e., a non-vanishing and negative cosmological constant, it is possible to obtain the real scalar field outside the horizon and the scalar field does not become ghostlike with resorting to eq.~(\ref{yueshu1}). Furthermore, we notice that the thermodynamic behavior of this class of Horndeski black holes with or without the negative cosmological constant is similar to that of the Schwarzschild-AdS  black hole. In fact, the negative cosmological constant $\Lambda$ not only plays the role of the thermodynamic pressure, but also gives the constraint eq.~(\ref{yueshu1}) to ensure the reality of the scalar field outside the horizon. Moreover, in the light of similar behaviors of the three thermodynamic quantities between the cases of different $\eta$ but fixed $\Lambda$ and the cases of different $\Lambda$ but fixed $\eta$, as shown in Figures \ref{tu1}, \ref{tu2} and \ref{tu3}, we can conclude that the newly introduced  intensive thermodynamic variable eq.~(\ref{new}) plays the similar role to that of thermodynamic pressure. At the end of this subsection, it is necessary to mention that  the temperature has one local minimum and the heat capacity at constant pressure undergoes only one divergence, see Figures \ref{tu1} and \ref{tu2}. These behaviors imply that there exists only one phase transition for the Horndeski black holes without charge. In the next subsection, we shall point out that the Horndeski black holes with the charge hair present rich thermodynamic behaviors and critical phenomena.

\subsection{Scenario with charge}
For the class of Horndeski black holes with an electric field, the analytic solution reads~\cite{ACCE,XHHC}
\begin{eqnarray}
f(r)&=&\frac{\alpha r^2}{3\eta}-\frac{2M}{r}+\frac{3\alpha+\Lambda \eta}{\alpha-\Lambda \eta}+\left(\frac{\alpha+\Lambda \eta+\frac{\alpha^2 q^2}{4\eta}}{\alpha-\Lambda \eta}\right)^2\,\frac{\tan^{-1}\left(\sqrt{\frac{\alpha}{\eta}}r\right)}{\sqrt{\frac{\alpha}{\eta}}r}\nonumber \\
& &+\frac{\alpha^2 q^2}{(\alpha-\Lambda \eta)^2 r^2}-\frac{\alpha^2 q^4}{48(\alpha-\Lambda \eta)^2 r^4}+\frac{\alpha^3 q^4}{16\eta(\alpha-\Lambda \eta)^2 r^2},\\
\vspace{0.8cm}
g(r)&=&\frac{\alpha^2[4(\alpha-\Lambda \eta) r^4+8\eta r^2-\eta q^2]^2}{16r^4(\alpha-\Lambda \eta)^2(\alpha r^2+\eta)^2 f(r)},\\
\vspace{0.8cm}
\psi^2(r)&=&-\frac{\alpha^2[4(\alpha+\Lambda \eta)r^4+\eta q^2][4(\alpha-\Lambda \eta) r^4+8\eta r^2-\eta q^2]^2}{32\eta r^6(\alpha-\Lambda \eta)^2 (\alpha r^2+\eta)^3 f(r)},\\
\vspace{0.8cm}
\Psi(r)&=&\Psi_0+\frac14 \frac{q\sqrt{\alpha}}{\eta^{\frac32}}\left[\frac{4\eta(\alpha+\Lambda\eta)+\alpha^2 q^2}{\alpha-\Lambda\eta}\right]\tan^{-1}\left(\sqrt{\frac{\alpha}{\eta}}r\right)\nonumber \\
& &+\frac{\alpha q(\alpha q^2+8\eta)}{4\eta r(\alpha-\Lambda\eta)}-\frac{\alpha q^3}{12r^3(\alpha-\Lambda\eta)}, \label{elepo}
\end{eqnarray}
where $\Psi_0$ is an integration constant. In order to obtain a real scalar field outside the horizon, one needs to impose the following constraint of parameters~\cite{ACCE},
\begin{equation}
4(\alpha+\Lambda \eta)r^4+\eta q^2 <0,  \label{yueshu2}
\end{equation}
whose derivation is similar to that of the constraint eq.(\ref{yueshu1}), see the analysis in the above subsection.

From the point of view of thermodynamics, the thermodynamic enthalpy can be written as a function of the horizon radius $r_h$,
\begin{eqnarray}
M&=&\frac{\alpha r_h^3}{6\eta}+\frac{(3\alpha+\Lambda \eta)r_h}{2(\alpha-\Lambda \eta)}+\frac12\sqrt{\frac{\eta}{\alpha}}\left(\frac{\alpha+\Lambda \eta+\frac{\alpha^2 q^2}{4\eta}}{\alpha-\Lambda \eta}\right)^2\tan^{-1}\left(\sqrt{\frac{\alpha}{\eta}}r_h\right)\nonumber\\
& &+\frac{\alpha^2 q^2}{2(\alpha-\Lambda \eta)^2 r_h}-\frac{\alpha^2 q^4}{96(\alpha-\Lambda \eta)^2 r_h^3}+\frac{\alpha^3 q^4}{32\eta(\alpha-\Lambda \eta)^2 r_h}. \label{enth2}
\end{eqnarray}
There are four pairs of thermodynamic variables and in each pair the two variables are conjugate to each other. By following the calculations of eqs.~(\ref{temp1}), (\ref{case1V}), and (\ref{case1Theta}), we derive the first three pairs using eq.~(\ref{enth2}).  The first pair consists of the Bekenstein-Hawking entropy eq.~(\ref{entro}) and the thermodynamic temperature,
\begin{eqnarray}
T_h&=&\frac{1}{4\pi r_h}\left\{\frac{\alpha r_h^2}{\eta}+\frac{3\alpha+\Lambda \eta}{\alpha-\Lambda \eta}+\frac{\alpha^2 q^4}{16r_h^4(\alpha-\Lambda\eta)^2}-\frac{\alpha^2 q^2}{r_h^2(\alpha-\Lambda\eta)^2}\right.\nonumber\\
& &\left.-\frac{\alpha^3 q^4}{16\eta r_h^2(\alpha-\Lambda\eta)^2}+\frac{[\alpha^2 q^2+4\eta(\alpha+\Lambda \eta)]^2}{16\eta(\alpha r_h^2+\eta)(\alpha-\Lambda \eta)^2}\right\}. \label{temp2}
\end{eqnarray}
The second pair contains the thermodynamic pressure eq.~(\ref{press}) and the thermodynamic volume,
\begin{eqnarray}
V&=&-\frac{16\pi \alpha \eta}{(\alpha-\Lambda \eta)^3}\left\{(\alpha-\Lambda \eta)r_h+\frac{\alpha q^4(3\alpha r_h^2-\eta)}{96\eta r_h^3}+\frac{\alpha q^2}{2r_h}\right.\nonumber\\
& &\left.+\sqrt{\frac{\eta}{\alpha}}\frac{(\alpha q^2+8\eta)[\alpha^2 q^2+4\eta(\alpha+\Lambda \eta)]}{32\eta^2}\tan^{-1}\left(\sqrt{\frac{\alpha}{\eta}}r_h\right)\right\}.
\end{eqnarray}
The last pair is composed of the coupling strength eq.~(\ref{new}) as a new intensive thermodynamic variable and its conjugate extensive variable,
\begin{eqnarray}
\Theta&=&\frac{4\pi r_h^3}{3}+\frac{\pi\alpha\Lambda\eta^2 q^4}{6r_h^3(\alpha-\Lambda\eta)^3}-\frac{8\pi\alpha\Lambda\eta^2 q^2}{r_h(\alpha-\Lambda\eta)^3}-\frac{16\pi\Lambda\eta^2 r_h}{(\alpha-\Lambda\eta)^2}+\frac{\pi\alpha^2 q^4(1-2\Lambda q)}{4r_h(\alpha-\Lambda\eta)^3}\nonumber\\
& &+\frac{2\pi \eta^2 r_h}{\alpha(\alpha r_h^2+\eta)}\left(\frac{\alpha+\Lambda\eta+\frac{\alpha^2 q^2}{4\eta}}{\alpha-\Lambda \eta}\right)^2-2\pi \left(\frac{\eta}{\alpha}\right)^{\frac32}\left(\frac{\alpha+\Lambda \eta+\frac{\alpha^2 q^2}{4\eta}}{\alpha-\Lambda \eta}\right)^2\tan^{-1}\left(\sqrt{\frac{\alpha}{\eta}}r_h\right)\nonumber \\
& &-2\pi\sqrt{\frac{\eta}{\alpha}}\,\frac{\alpha+\Lambda \eta+\frac{\alpha^2 q^2}{4\eta}}{(\alpha-\Lambda \eta)^3}\left(8\Lambda\eta^2+2\alpha\Lambda\eta q^2-\alpha^2 q^3\right)\tan^{-1}\left(\sqrt{\frac{\alpha}{\eta}}r_h\right).
\end{eqnarray}
In addition, due to the presence of an electric filed, the charge $Q$ of black holes reads
\begin{equation}
Q=\frac{\alpha q}{\alpha-\Lambda\eta},
\end{equation}
and its conjugate intensive variable, i.e., the electric potential $\Phi$ can be obtained,
\begin{eqnarray}
\Phi&=&\left(\frac{\partial M}{\partial Q}\right)_{S,P,\Pi}\nonumber\\
&=&\frac14 \frac{q\sqrt{\alpha}}{\eta^{\frac32}}\left(\frac{4\eta(\alpha+\Lambda\eta)+\alpha^2 q^2}{\alpha-\Lambda\eta}\right)\tan^{-1}\left(\sqrt{\frac{\alpha}{\eta}}r_h\right)\nonumber\\
& &+\frac{\alpha q(\alpha q^2+8\eta)}{4\eta r_h(\alpha-\Lambda\eta)}-\frac{\alpha q^3}{12r_h^3(\alpha-\Lambda\eta)}.\label{elepo11}
\end{eqnarray}
We notice that eq.~(\ref{elepo11}) has a good agreement with eq.~(\ref{elepo}) under the condition of the vanishing integration constant $\Psi_0=0$. Eq.~(\ref{elepo}) was obtained by solving the Maxwell equation of motion eq.~(\ref{emu}), i.e., $\nabla_{\mu}F^{\mu\nu}=0$, while eq.~(\ref{elepo11}) is derived by us through thermodynamic relations. The consistency of the two equations, eqs.~(\ref{elepo}) and  (\ref{elepo11}), shows that the thermodynamic method we have adopted is reasonable.

Hence, we give the first law of thermodynamics and the generalized Smarr relation as follows,
\begin{eqnarray}
\text{d}M&=& T_h \text{d}S+V\text{d}P+\Phi \text{d}Q+\Theta \text{d}\Pi,\nonumber\\
\vspace{0.8cm}
M&=& 2T_h S-2PV+\Phi Q-2\Pi\Theta.
\label{law3}
\end{eqnarray}

With the help of eqs.~(\ref{entro}), (\ref{enth2}), and (\ref{temp2}), we can derive the heat capacity at constant pressure,
\begin{equation}
C_p \equiv \left(\frac{\partial M}{\partial T_h}\right)_P=\frac{\partial M}{\partial r_h}\left(\frac{\partial T_h}{\partial r_h}\right)^{-1}, \label{capac2}
\end{equation}
where the factors of the numerator and denominator of eq.~(\ref{capac2}) can be calculated to be, respectively,
\begin{eqnarray*}
\frac{\partial M}{\partial r_h} &=& \frac{\alpha r_h^2}{2\eta}+\frac{\eta(\alpha+\Lambda\eta)^2}{(\alpha r_h^2+\eta)(\alpha-\Lambda\eta)^2}+\frac{3\alpha+\Lambda\eta}{\alpha-\Lambda\eta}+\frac{\alpha^2 q^4}{32r_h^4(\alpha-\Lambda\eta)^2}\\
& &-\frac{\alpha^2 q^2}{2r_h^2(\alpha-\Lambda\eta)^2}-\frac{\alpha^3 q^4}{32\eta r_h^2(\alpha-\Lambda\eta)^2},\\
\vspace{0.4cm}
\frac{\partial T_h}{\partial r_h}&=&-\frac{1}{4\pi r_h^2}\left[-\frac{\alpha r_h^2}{\eta}+\frac{3\alpha+\Lambda\eta}{\alpha-\Lambda\eta}+\frac{\eta\left(\alpha+\Lambda\eta+\frac{\alpha^2 q^2}{4\eta}\right)^2}{(\alpha r_h^2+\eta)(\alpha-\Lambda\eta)^2}-\frac{3\alpha^2 q^2}{r_h^2(\alpha-\Lambda\eta)^2}\right.\\
& &\left.+\frac{5\alpha^2 q^4}{16r_h^4(\alpha-\Lambda\eta)^2}-\frac{3\alpha^3 q^4}{16\eta r_h^2(\alpha-\Lambda\eta)^2}+\frac{2\alpha\eta r_h^2\left(\alpha+\Lambda\eta+\frac{\alpha^2 q^2}{4\eta}\right)^2}{(\alpha r_h^2+\eta)^2(\alpha-\Lambda\eta)^2}\right],
\end{eqnarray*}
and the Gibbs free energy,
\begin{eqnarray}
G &\equiv& M-T_h S \nonumber\\
&=&-\frac{\alpha r_h^3}{12\eta}-\frac{\eta r_h\left(\alpha+\Lambda \eta+\frac{\alpha^2 q^2}{4\eta}\right)^2}{4(\alpha r_h^2+\eta)(\alpha-\Lambda \eta)^2}+\frac12\sqrt{\frac{\eta}{\alpha}}\left(\frac{\alpha+\Lambda \eta+\frac{\alpha^2 q^2}{4\eta}}{\alpha-\Lambda \eta}\right)^2 \tan^{-1}\left(\sqrt{\frac{\alpha}{\eta}}r_h\right)\nonumber\\
& &+\frac{(3\alpha+\Lambda \eta)r_h}{4(\alpha-\Lambda \eta)}-\frac{5\alpha^2 q^4}{192r_h^3(\alpha-\Lambda\eta)^2}+\frac{3\alpha^2 q^2}{4r_h(\alpha-\Lambda\eta)^2}+\frac{3\alpha^3 q^4}{64\eta r_h(\alpha-\Lambda\eta)^2}.  \label{gfree2}
\end{eqnarray}

\begin{figure}
\begin{center}
  \begin{tabular}{cc}
    \includegraphics[width=70mm]{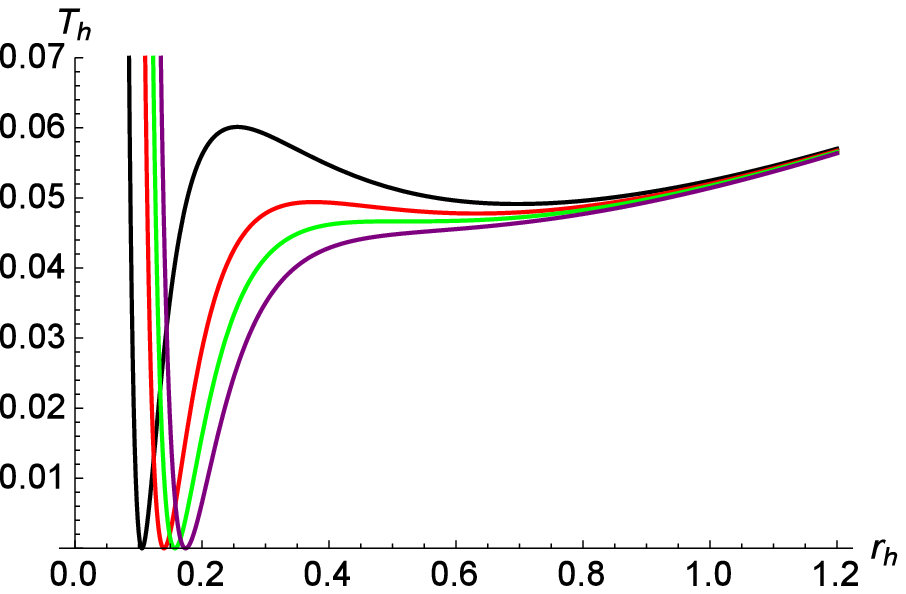} &
    \includegraphics[width=70mm]{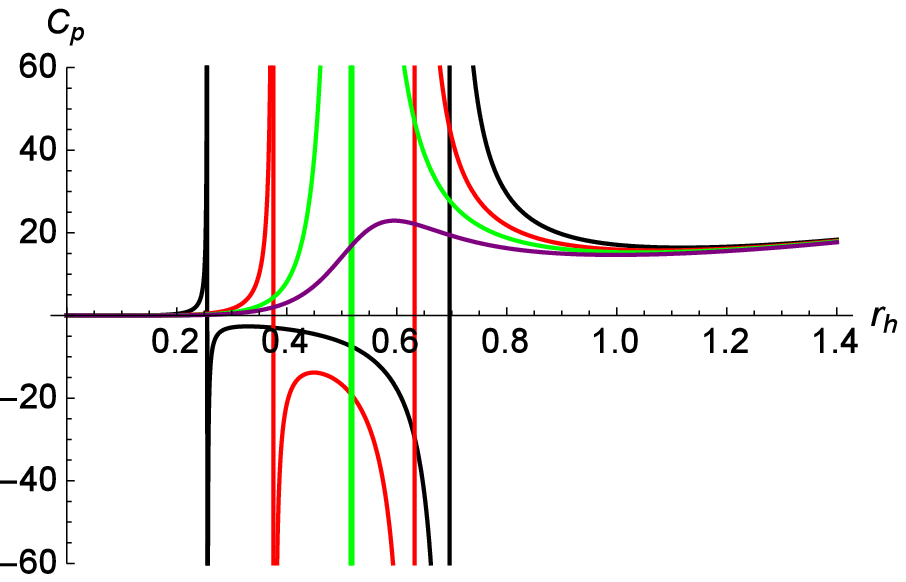}
  \end{tabular}
\end{center}
\caption{When $\alpha=1$, $\eta=2$, and $\Lambda=-1.5$, plots of the relations of $T_h$ and $C_p$ with respect to $r_h$ for $q=0.3$  (\text{Black}), $q=0.4$ (\text{Red}), $q_c=0.45049$ (\text{Green}), and $q=0.5$ (\text{Purple}), respectively.}
\label{tu4}
\end{figure}

\begin{figure}
\begin{center}
  \begin{tabular}{cc}
    \includegraphics[width=70mm]{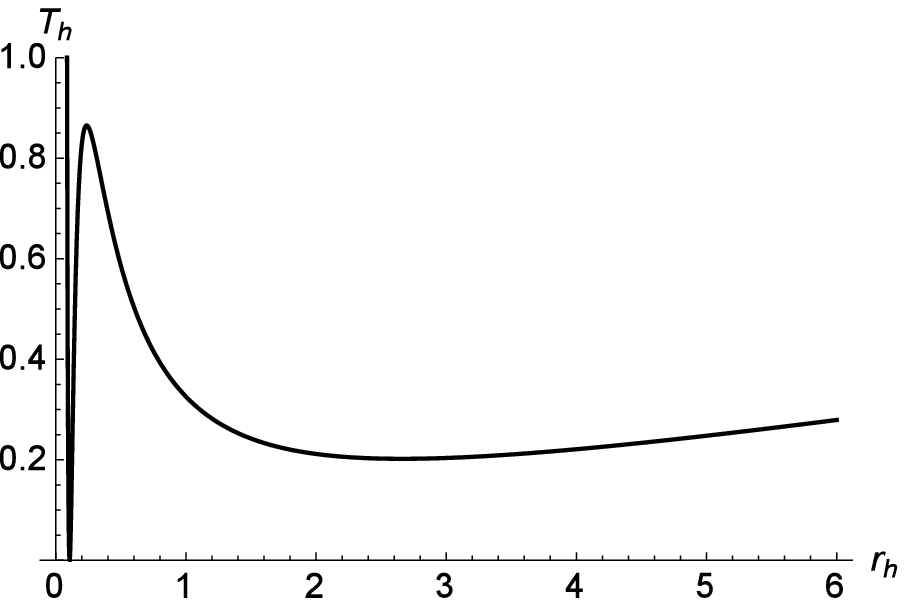} &
    \includegraphics[width=70mm]{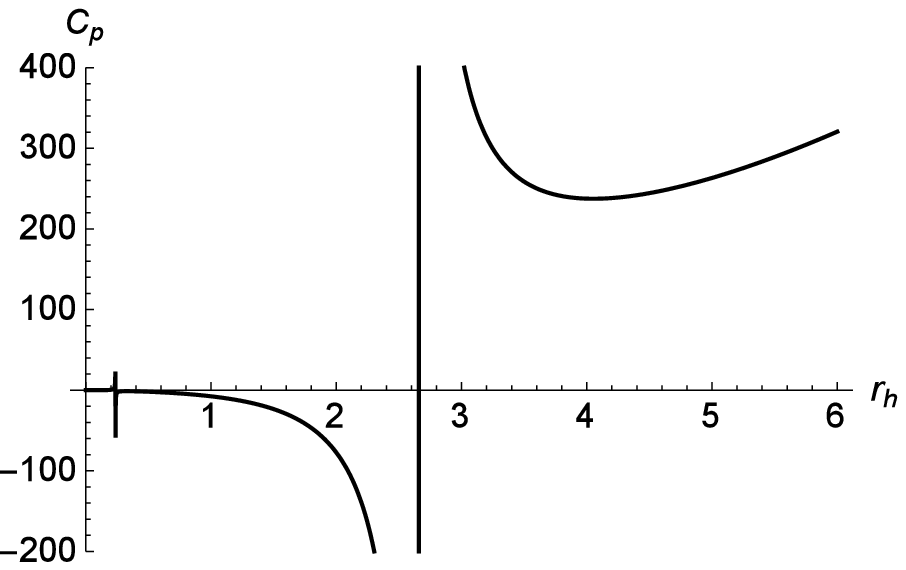} \\
    \includegraphics[width=70mm]{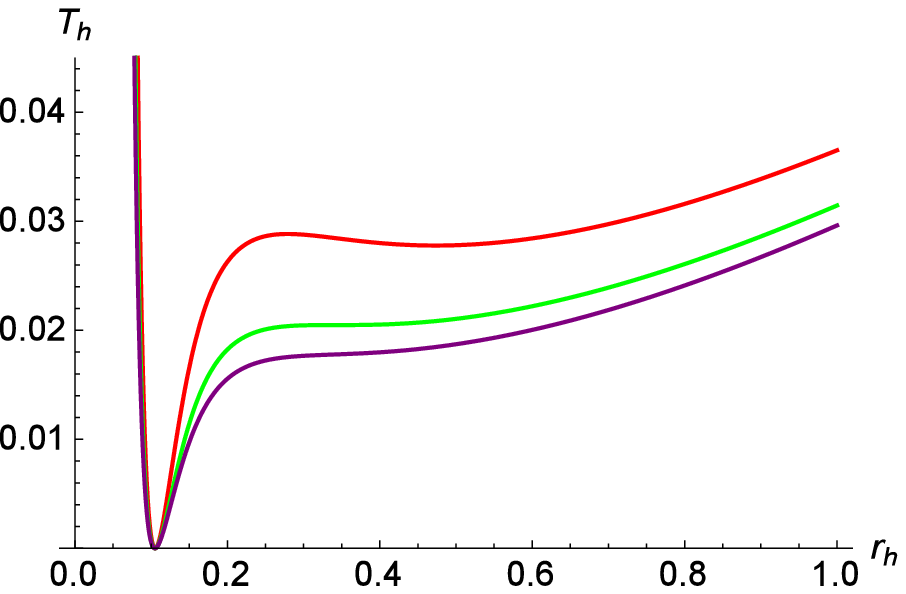} &
    \includegraphics[width=70mm]{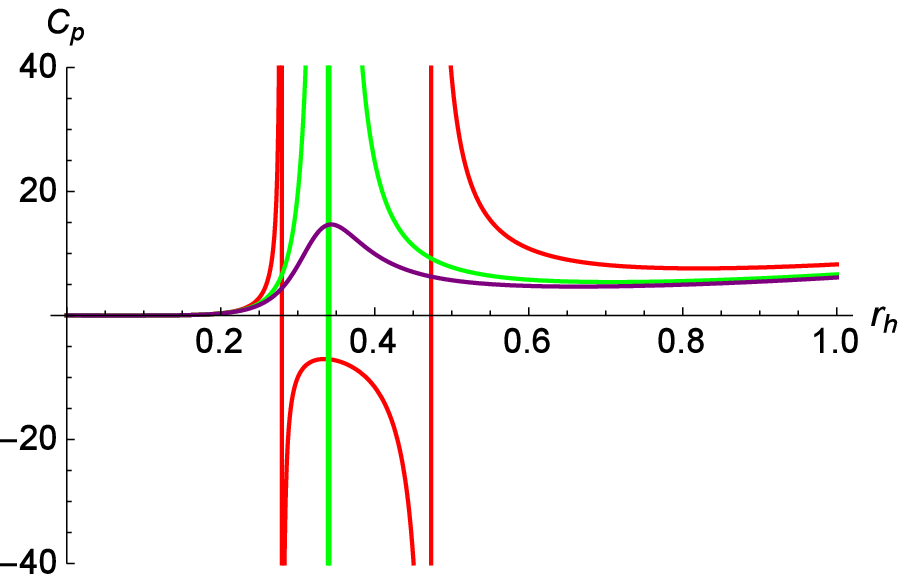}
  \end{tabular}
\end{center}
\caption{When $\alpha=1$, $\eta=2$, and $q=0.3$, plots of the relations of $T_h$ and $C_p$ with respect to $r_h$ for $\Lambda=0$ (\text{Black}), $\Lambda=-2.5$ (\text{Red}), $\Lambda_c=-3.1631$ (\text{Green}), and $\Lambda=-3.5$ (\text{Purple}), respectively.}
\label{tu5}
\end{figure}

\begin{figure}
\begin{center}
  \begin{tabular}{cc}
    \includegraphics[width=70mm]{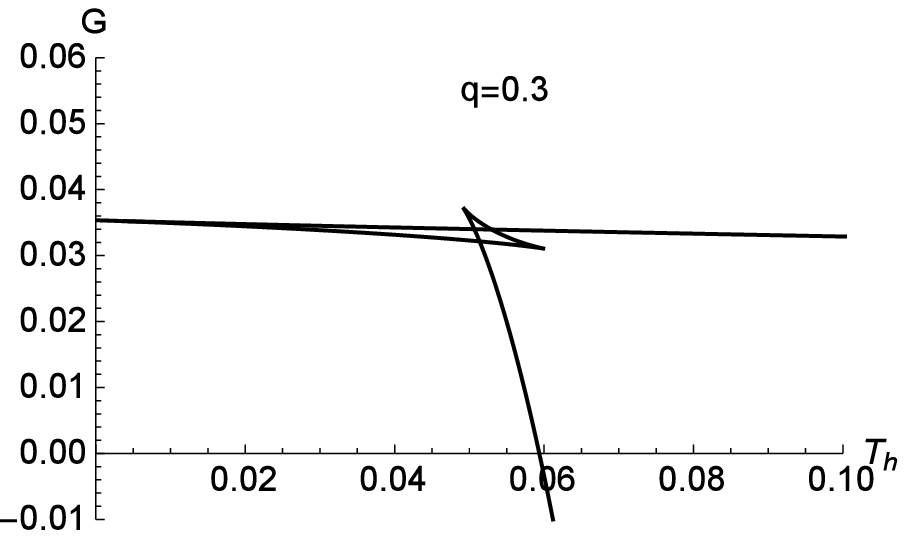} &
    \includegraphics[width=70mm]{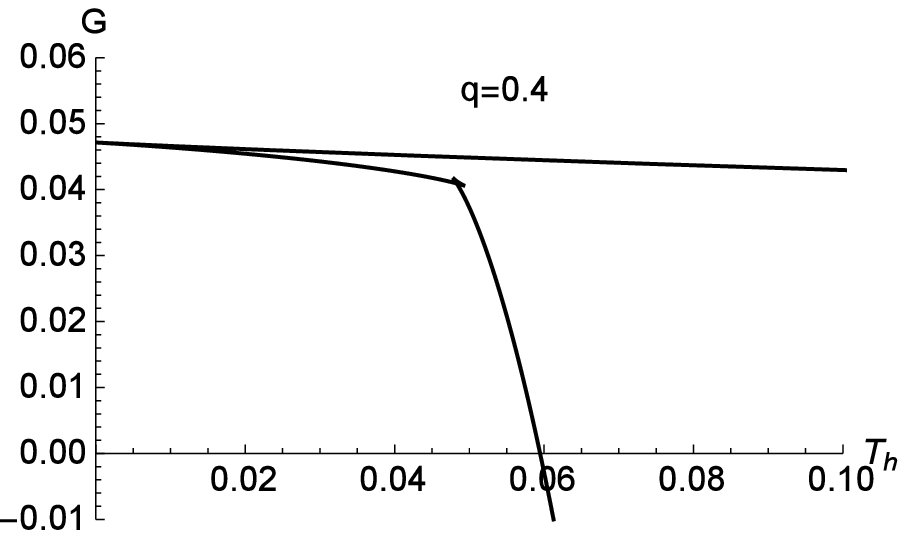} \\
    \includegraphics[width=70mm]{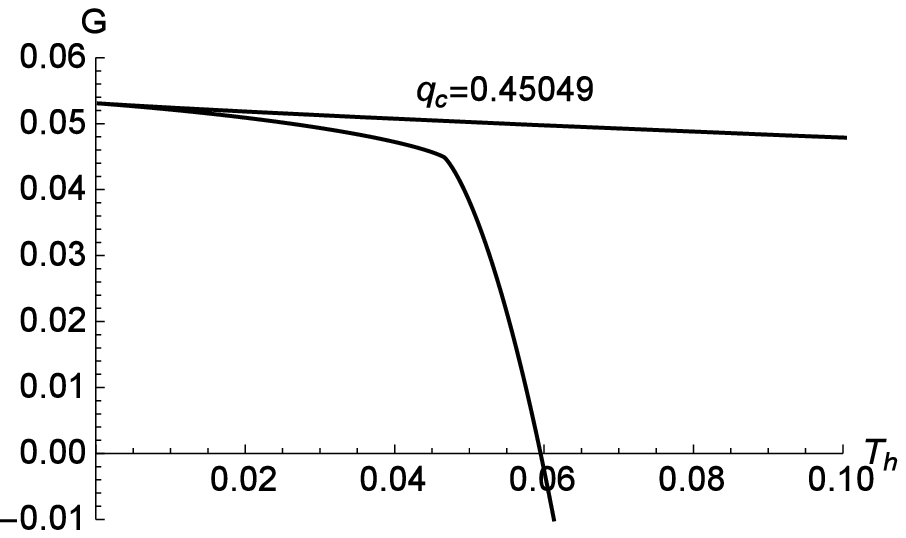} &
    \includegraphics[width=70mm]{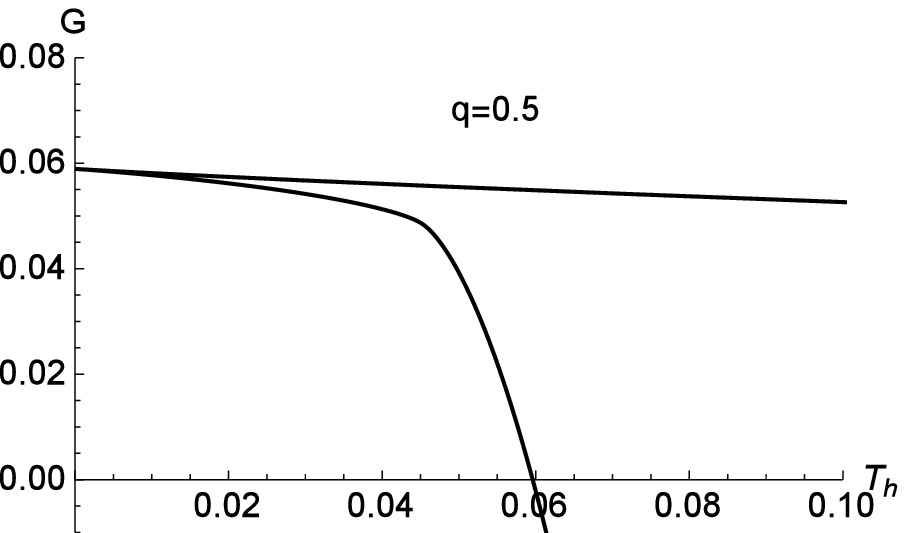}
  \end{tabular}
\end{center}
\caption{When $\alpha=1$, $\eta=2$, and $\Lambda=-1.5$, plots of the relation of $G$ with respect to $T_h$ for different values of $q$.}
\label{tu6}
\end{figure}

\begin{figure}
\begin{center}
  \begin{tabular}{cc}
    \includegraphics[width=70mm]{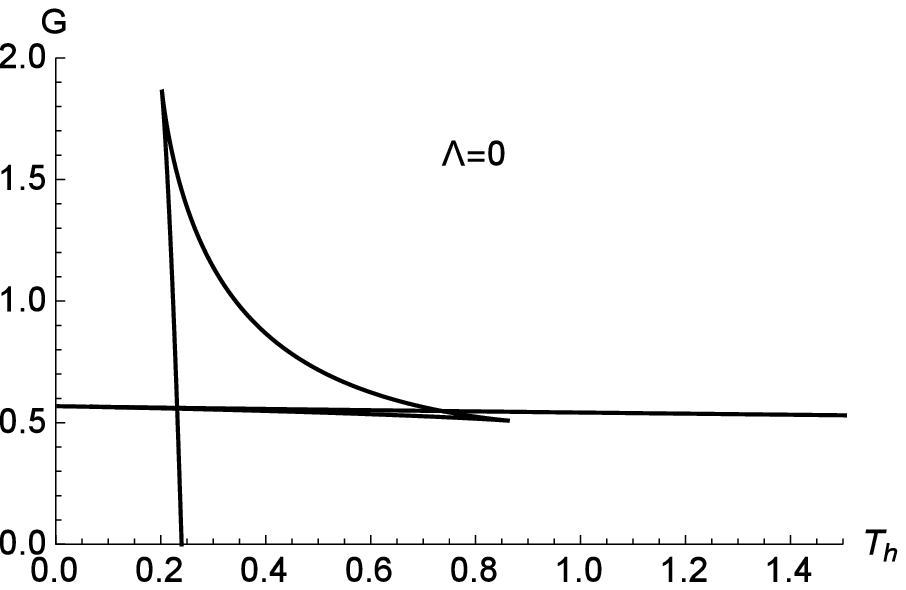} &
    \includegraphics[width=70mm]{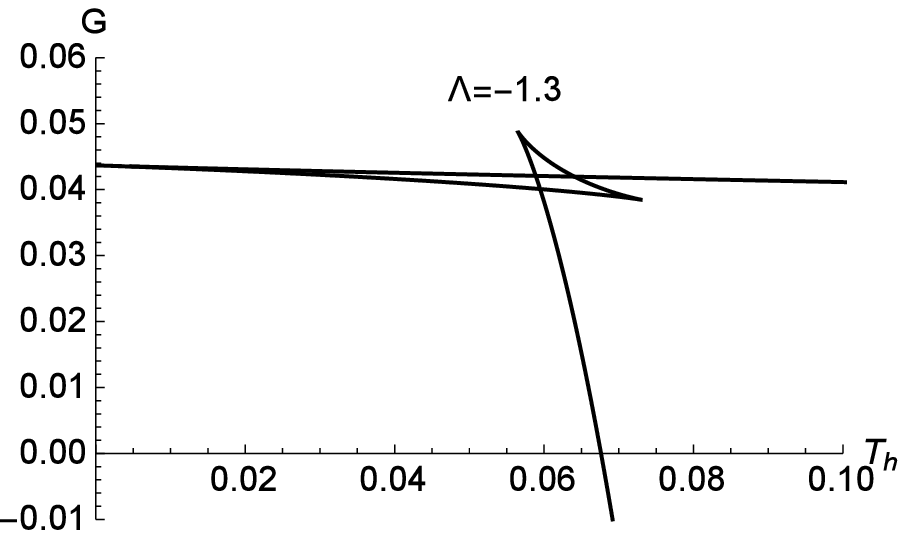} \\
    \includegraphics[width=70mm]{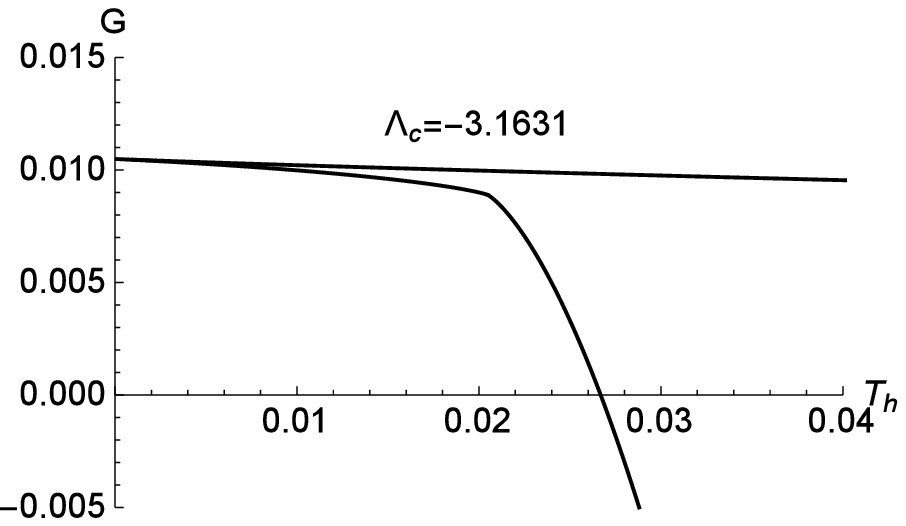} &
    \includegraphics[width=70mm]{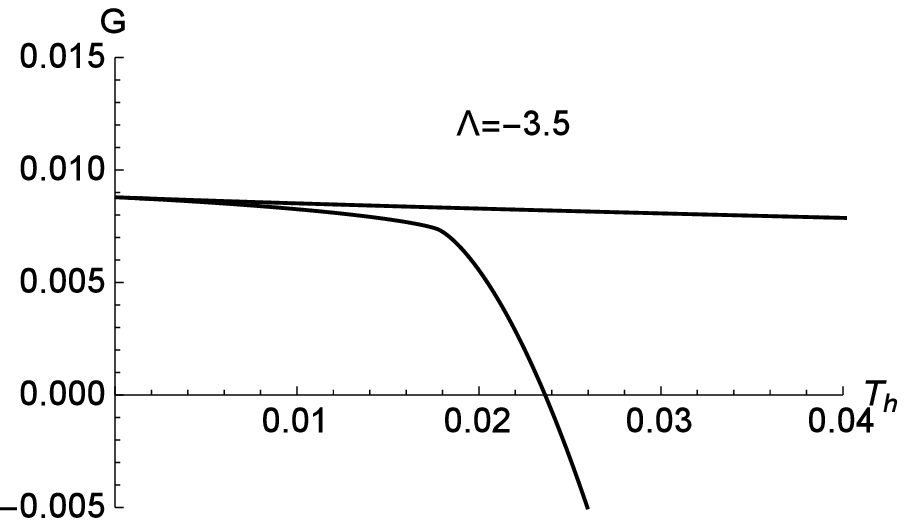}
  \end{tabular}
\end{center}
\caption{When $\alpha=1$, $\eta=2$, and $q=0.3$, plots of the relation of $G$ with respect to $T_h$ for different values of $\Lambda$.}
\label{tu7}
\end{figure}

We plot the thermodynamic temperature eq.~(\ref{temp2}), the heat capacity at constant pressure eq.~(\ref{capac2}), and the Gibbs free energy eq.~(\ref{gfree2}) in Figures \ref{tu4}, \ref{tu5}, \ref{tu6}, and \ref{tu7}. These figures show rich thermodynamic behaviors and critical phenomena if we compare to the case of no charge. In Figure \ref{tu4}, for fixed values of $\alpha$, $\eta$, and $\Lambda$, the thermodynamic temperature presents one local maximum and one local minimum, and the heat capacity at constant pressure undergoes two times of divergence for different values of $q$. As is known, black holes are locally stable for $C_p > 0$, while unstable for $C_p <0$. The behavior of the heat capacity at constant pressure depicted by Figure \ref{tu4} indicates that the Horndeski black holes undergo two phase transitions: the first phase transition happens from a locally stable state to a locally unstable one at the local maximum temperature, and the other phase transition occurs from a locally unstable state to a locally stable one at the local minimum temperature. When $q$ is larger than its critical value $q_c=0.45049$, the temperature has no extrema and the heat capacity at constant pressure has no divergences. In addition, the characteristic swallowtail behavior of the Gibbs free energy disappears once $q>q_c=0.45049$, as shown in Figure \ref{tu6}. All of these phenomena imply that no phase transitions happen when the charge parameter $q$ exceeds its critical value $q_c$. On the other hand, the similar critical phenomenon appears for different values of $\Lambda$ but fixed values of $\alpha$, $\eta$, and $q$, as shown in Figures \ref{tu5} and \ref{tu7}. When the cosmological parameter $\Lambda$ is less than its critical value $\Lambda_c=-3.1631$, no phase transitions occur.  Furthermore, we observe that the thermodynamic temperature goes to zero at a very small horizon radius. It is the electric charge that provides negative contributions in eq.~(\ref{temp2}), which makes it possible that the thermodynamic temperature vanishes.

\section{Conclusion}\label{sec4}
Based on the well-accepted consideration~\cite{CEJM,DSJT,BPD2,BPD,KM}, namely the cosmological constant as the thermodynamic pressure and the mass of black holes as thermodynamic enthalpy, we revisit thermodynamic properties of a new class of Horndeski black holes whose action contains a non-minimal kinetic coupling of one massless real scalar and the Einstein tensor. We resort to a new intensive thermodynamic variable, see eq.~(\ref{new}), which originates from the coupling strength $\eta$ with the dimension of length square, and thus deduce both the generalized first law of thermodynamics and the generalized Smarr relation, see eqs.~(\ref{law1}), (\ref{law2}), and (\ref{law3}). By calculation of some thermodynamic quantities, such as the thermodynamic temperature, the heat capacity at constant pressure, and the Gibbs free energy, our result indicates that this class of Horndeski black holes presents rich thermodynamic behaviors and critical phenomena. Especially in the case of the presence of an electric field, the black holes undergo two times of phase transitions: the first phase transition happens from a locally stable state to a locally unstable one, and the other phase transition occurs from a locally unstable state to a locally stable one. Once the charge parameter $q$ exceeds its critical value $q_c$, or the cosmological parameter $\Lambda$ does not exceeds its critical value $\Lambda_c$, no phase transitions happen and the black holes are stable. As a by-product, we indicate that the behavior of the coupling strength acts as that of the thermodynamic pressure, as shown in Figures \ref{tu1}, \ref{tu2}, and \ref{tu3}.

\section*{Acknowledgments}
Y-GM would like to thank H. Nicolai of Max-Planck-Institut f\"ur Gravitationsphysik (Albert-Einstein-Institut) for kind hospitality. This work was supported in part by the National Natural
Science Foundation of China under grant No.11675081. At last, the authors would like to thank the anonymous referee for the helpful comment that indeed greatly improves this work.

\end{document}